\documentclass{sigchi}

\CopyrightYear{2020}
\setcopyright{acmlicensed}
\doi{https://doi.org/10.1145/3313831.XXXXXXX}
\isbn{978-1-4503-6708-0/20/04}
\conferenceinfo{CHI'20,}{April  25--30, 2020, Honolulu, HI, USA}
\acmPrice{\$15.00}

\usepackage{balance}       
\usepackage{graphics}      
\usepackage[T1]{fontenc}   
\usepackage{txfonts}
\usepackage{mathptmx}
\usepackage[pdflang={en-US},pdftex]{hyperref}
\usepackage{color}
\usepackage{booktabs}
\usepackage{textcomp}
\usepackage{enumitem}

\usepackage{microtype}        
\usepackage{ccicons}          

\usepackage{todonotes}
\usepackage{scalerel}
\usepackage{wrapfig}
\newcommand{\cmo}{\textcolor[rgb]{0, 0, 0}}
\DeclareRobustCommand{\inlinefig}[1]{%
\begingroup
\setbox0=\hbox{\includegraphics[height=1.1em]{#1}}%
\parbox{\wd0}{\box0}\endgroup
}

\def\plaintitle{SIGCHI Conference Proceedings Format} 

\def\emptyauthor{}
\def\plainkeywords{Visualization in Augmented Reality; Augmented Static Visualization; Data Visualization Authoring.}

\makeatletter
\def\url@leostyle{%
  \@ifundefined{selectfont}{
    \def\UrlFont{\sf}
  }{
    \def\UrlFont{\small\bf\ttfamily}
  }}
\makeatother
\def\pprw{8.5in}
\def\pprh{11in}

\definecolor{linkColor}{RGB}{6,125,233}
\hypersetup{%
  pdftitle={\plaintitle},
  pdfauthor={\emptyauthor},
  pdfkeywords={\plainkeywords},
  pdfdisplaydoctitle=true, 
  bookmarksnumbered,
  pdfstartview={FitH},
  colorlinks,
  citecolor=black,
  filecolor=black,
  linkcolor=black,
  urlcolor=linkColor,
  breaklinks=true,
  hypertexnames=false
}

\begin{document}

\newcommand{\ben}[1]{\textcolor{red}{\textit{ben:}#1}}
\newcommand{\qianwen}[1]{\textcolor{blue}{\textit{wqw:}#1}}
\newcommand{\tool}{PapARVis Designer}
\newcommand{\feata}{\textit{AR-preview}}
\newcommand{\featb}{\textit{Validator}}
\newcommand{\keepvalues}{%
  \edef\restorevalues{%
    \intextsep=\the\intextsep
    \columnsep=\the\columnsep
  }%
}

\title{Augmenting Static Visualizations with \tool}
\author{
Chen Zhu-Tian\textsuperscript{1}, 
Wai Tong\textsuperscript{1},
Qianwen Wang\textsuperscript{1},
Benjamin Bach\textsuperscript{2},
Huamin Qu\textsuperscript{1}\\
\\
{\large\normalfont 
\textsuperscript{1} Hong Kong University of Science and Technology, Hong Kong,
\textsuperscript{2} University of Edinburgh, UK
}
\\
{\large\normalfont 
\{zhutian.chen, wtong, qwangbb\}@connect.ust.uk,
bbach@ed.ac.uk,
huamin@ust.hk}\\
}

\teaser{ 
    \centering
    \includegraphics[width=2.1\columnwidth]{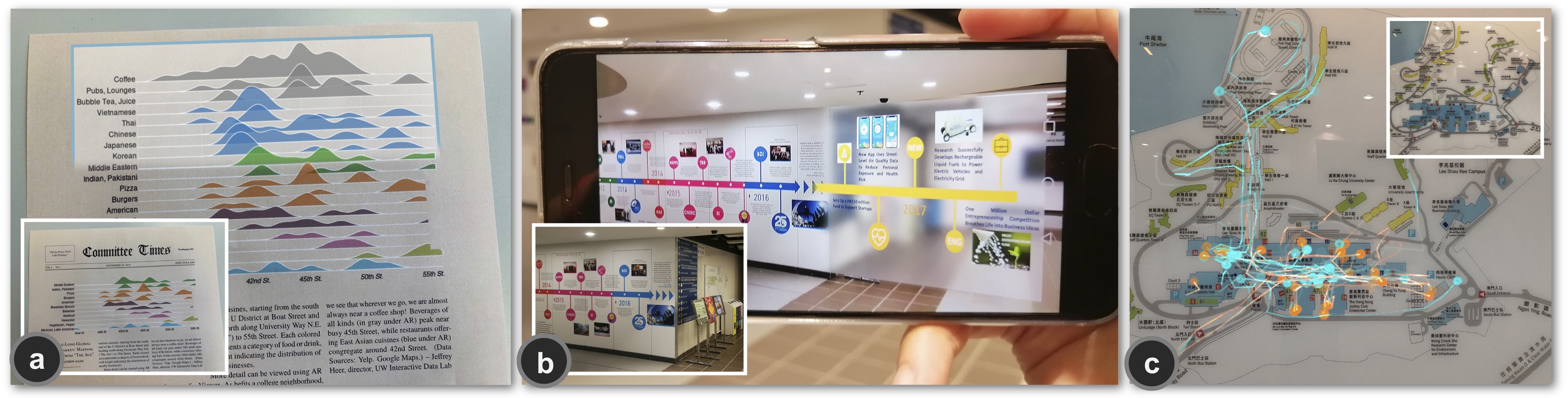}
    \caption{Augmenting static visualizations can leverage the best of both physical and digital worlds:
       a) a data journalism uses an augmented static visualization to extend the space of the newspaper that is limited by the banner; 
       b) a designer uses AR to update the outdated wall-sized timeline without recreating it;
       c) a tourist overlays the trajectories data on a public map in AR to see his/her moving pattern.
    }
    \label{fig:outdated-tl}
    \vspace{-1em}
} 
\maketitle

\begin{abstract}
This paper presents an authoring environment for augmenting static visualizations with virtual content in augmented reality. 
Augmenting static visualizations can leverage the best of both physical and digital worlds, but its creation currently involves different tools and devices, without any means to explicitly design and debug both static and virtual content simultaneously. 
To address these issues, we design an environment that seamlessly integrates all steps of a design and deployment workflow through its main features: \textit{i)} an extension to Vega, \textit{ii)} a preview, and \textit{iii)} debug hints that facilitate valid combinations of static and augmented content. 
We inform our design through a design space with four ways to augment static visualizations.
We demonstrate the expressiveness of our tool through examples, including books, posters, projections, wall-sized visualizations.
A user study shows high user satisfaction of our environment and confirms that participants can create augmented visualizations in an average of 4.63 minutes.

\end{abstract}

\ccsdesc[500]{Human-centered computing~Visualization systems and tools}
\keywords{\plainkeywords}

\printccsdesc

\newpage
\section{Introduction}

Data visualizations---in the form of posters, newspapers, scientific reports, on public displays as well as slideshows and mobile screens---become increasingly more widespread. 
In most of these cases, visualizations are printed on paper, 
which, due to its extremely low-tech and tangibility, 
makes it very simple to create, distribute, view, and engage with the visualization content: 
they can be viewed in-situ without requiring any specific hardware;
viewers can freely engage with them through touch, pen and annotation,
and sharing thoughts and discussions in collaborative settings.
\cmo{However, a drawback of paper visualizations is that these visualizations are \textit{static}, 
\textit{i.e.}, the displayed information is limited in both \textit{space} and \textit{time}.}

\cmo{Augmented Reality (AR)
allows for dynamic and interactive content, 
by adding an extra dynamic display layer onto an existing visualization. 
This combination can complement the static visualizations through extra data (\autoref{fig:outdated-tl}a), update outdated data (\autoref{fig:outdated-tl}b), highlight data (\autoref{fig:outdated-tl}c),
show details,
protect privacy, 
provide 3D content and interactivity~\cite{visar}, \emph{etc}.
We envision such kind of \emph{augmented static visualizations} 
being used for 
public display (\emph{e.g.}, information board, artworks in exhibitions), 
education (\emph{e.g.}, textbook), 
and creative products (\emph{e.g.}, giftcards), \emph{etc}.}

\cmo{At the same time, it is demanding to create combinations and hybrids of static and virtual AR-visualizations since a couple of criteria must be met to provide for a seamless and consistent integration. We can define these \textit{criteria of consistency} (C1-C3) as follows:
\begin{itemize}[noitemsep]
\item\textbf{C1: Graphical consistency} refers to equal graphical styles between static and virtual content, \emph{e.g.}, fonts, colors.
\item\textbf{C2: Readability} requires to provide for correct interpretation of visualizations across both media, \emph{e.g.}, aligning related (equal) graphical elements and labels, axes, and layouts; avoid visual clutter and overlap of graphical elements.
\item\textbf{C3: Validity} of visual encodings, 
\emph{e.g.}, values of the same visual variables mean the same data.
\end{itemize}
Support for maintaining these crucial characteristics is specifically more problematic when design iterations are required, 
\emph{e.g.}, trying different layouts, exploring the available space in AR. Without an integrated authoring approach, the design process is not just technically tedious, \emph{i.e.}, switching between authoring tools and environments, but can lead to inconsistent visualizations if a designer does not manage to manually account for consistency (C1-C3).}



 
To address C1-C3, 
this paper introduces \textit{\textbf{\tool{}}}, 
\cmo{an authoring environment to create augmented static visualizations.}
The design of \tool{} (Sec.\ref{sec:tool}) is based on a design space (Sec.\ref{sec:design_space}) 
that defines possible and valid ways of combining static and virtual content.
Similar to DXR~\cite{Sicat2019}, 
\tool{} 
delivers an extension to the Vega grammar~\cite{Satyanarayan2016},
leveraging its simplicity and expressiveness to specify 2D visualizations.
\cmo{\tool{} enables designers to create static and 2D virtual visualizations 
based on the same specification (C1,C3)
and to deploy everything with one click. 
To facilitate design and avoid invalid combinations, 
we included two bespoke features into \tool{}: 
first, \textbf{\feata{}} gives a preview of the augmented static visualization
that allows designers to view their designs on the desktop environment,
thus assuring \emph{Readability} and reducing switching between devices (C2);
second, \textbf{\featb{}} automatically validates a design
based on our design space 
and provides guidelines for correction to ensure 
consistency of the visual encodings
between the static and virtual visualizations (C3).}

To demonstrate the expressiveness of \tool{},
we provide an exemplary gallery of 
13 augmented static visualizations (\autoref{fig:outdated-tl} and \autoref{fig:gallery}), 
each varying in data type, 
chart type, 
media, 
and purpose. 
A controlled user study with 12 visualization designers but without
further experience in AR development suggests the overall usability of \tool{} and the two main features \feata{} and \featb{}. \tool{} is available at \url{https://github.com/PapARVis}.

\section{Related Work}

This section overviews prior research
on AR visualization,
augmenting physical documents,
and visualization tools.

\subsection{AR Visualization}

Willett et al.~\cite{Willett2017} introduced embedded data representations,
a conceptual framework to unify the research on visualization systems 
that connect to the physical world.
As an important method to blend digital data and the physical world,
AR has attracted the attention of the visualization community.
Benefits of visualizing data in AR have been reported by previous research.
First, for example,
AR can visualize data in the physical space 
to facilitate certain visual explorations and collaborative analysis.
Butscher et al.~\cite{Butscher2018} presented ART,
an AR collaborative analysis tool, 
and reported that ART could facilitate communication and coordination between collaborators, 
since the interaction is grounded in the physical world.
Besides, 
AR can augment physical objects with rich digital information,
enabling situated analytics.
Zhao et al.~\cite{sa-ego-centric} 
developed a mobile AR system to 
support on-site analysis of co-authoring patterns of researchers.
SiteLens~\cite{White2009} visualizes relevant data
in the 3D physical context for urban site visits.
ElSayed et al.~\cite{ElSayed2015} developed an AR system to help customers 
filtering, finding, and ranking products during shopping. 
Moreover, AR has been used to present 
visualizations for communication purposes~\cite{Chen2019}
as it has the potential to engage audiences better.
In summary, previous work shows that AR can 
connect virtual data to physical spaces,
augment real objects with rich information,
and even engage users.
We draw on this line of work 
and attempt to augment static visualizations 
with virtual visualizations in AR,
harnessing the best of both physical and digital media. 

\subsection{Augmenting Physical Documents}

A variety of AR systems 
(\emph{e.g.}, projector-based~\cite{Wellner91}, handled-based~\cite{BillinghurstKP01}, and HMD-based~\cite{Li2019}) have been proposed 
to augment physical documents 
by providing additional functionality and content.
For example, HoloDoc~\cite{Li2019},
a mixed reality system based on HoloLens,
augments physical artifacts to 
allow users to take notes and look up words.
Although these systems 
augment physical documents with rich virtual content,
they are not designed for augmenting static visualizations,
which requires a high level of 
information integrity,
visual style consistency,
and spatial alignment
between the static and virtual content.
Recently, initial explorations have been made 
to augment static visualizations:
Kim et al.~\cite{visar} presented VisAR,
an AR prototype system that 
provides interactive functions to static visualizations.
Differently,
we aim to use AR to extend static visualizations with additional data.

\subsection{Visualization Authoring Tools}

A number of authoring tools have been proposed
to facilitate the creation of visualizations 
in desktop and AR environments.
Prior work on creating desktop visualization
ranges from highly expressive programming toolkits
to easy-to-use WIMP UI tools.
Programming toolkits 
(\emph{e.g.}, D3~\cite{Bostock2011}, ProtoVis~\cite{protovis})
provide strong expressiveness and 
support flexible visualization designs
but require a high level of programming skills.
Instead,
WIMP UI systems 
(\emph{e.g.}, Lyra~\cite{lyra}, iVisDesigner~\cite{Ren2014}, and Voyager~\cite{Wongsuphasawat2016})
lower the barrier of creating visualizations 
by providing interactive functions
to allow users to design visualizations via drag and drop,
but compromising the flexibility.
To strike a balance between expressivity and simplicity,
Vega~\cite{Satyanarayan2016} and Vega-Lite~\cite{Satyanarayan2018}
enable users to define visualizations 
with a concise declarative specification in JSON format.
Overall, these systems 
focus on 2D visualizations on desktop platforms
but cannot create visualizations in AR.

On the other hand, tools have recently been proposed
for creating visualizations in AR environments.
For example,
MARVisT~\cite{Chen2019} is
a touch-based creativity support tool to assist general users in creating AR-based glyph visualizations.
Programmatic toolkits have presented to support more flexible visualizations;
DXR~\cite{Sicat2019} and IATK~\cite{Cordeil2019} are fast prototyping toolkits, 
that provide both programmatic 
and graphical user interfaces 
to help designers create immersive visualizations. These systems mainly focus on 3D visualizations, thus requiring knowledge of 3D computer graphics (\emph{e.g.}, shader, meshes, and 3D camera).

Most important, these tools do not envision extending any existing static visualization. First, none of these tools can create 
both static and virtual visualizations simultaneously,
thus requiring the designer to switch between tools frequently.
Second, 
none of these tools can help designer to ensure
the consistency of data, visual styles, and positions
between the static and virtual visualization.
Our \tool{} presents an authoring environment 
to create visualizations crossing between reality and virtuality.
\section{Design Space}
\label{sec:design_space}

This section discusses \cmo{the concept of} validity for augmented static visualizations, explores the design space \cmo{for spatially aligning static and virtual content}, and derives a set of design goals for our authoring environment.

\subsection{What kind of augmented static visualizations is valid?}
\label{ssec:model}
In this work,
we distinguish between these three terms:
\textit{i)} \textit{static visualizations} ($V_s$), which are static but can be in different media (\emph{e.g.}, be printed, projected, and displayed in digital screens),
\textit{ii)} \textit{virtual visualizations} ($V_v$),
which are displayed in AR 
(\emph{e.g.}, the virtual timeline in \autoref{fig:outdated-tl}b),
and \textit{iii)} \textit{augmented static visualizations} ($V_{ar}$),
which combine both static and virtual visualizations in AR.


\begin{figure}[h]
    \centering
    \includegraphics[width=0.7\columnwidth]{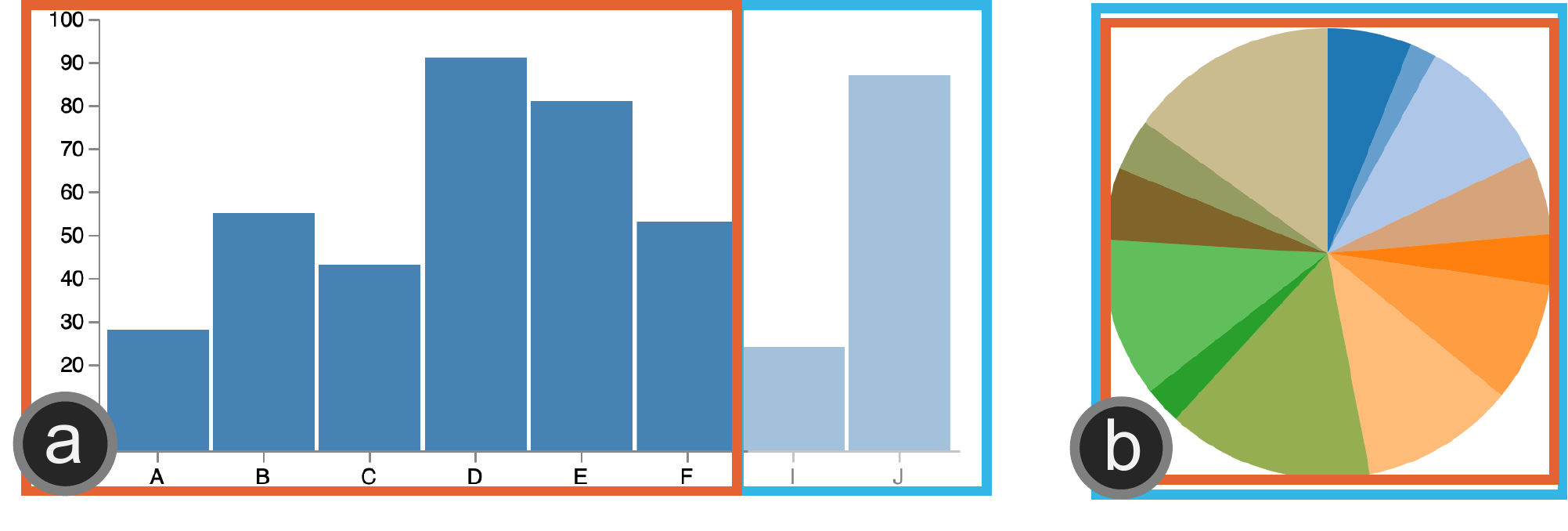}
    \caption{The validity of augmented static visualizations: 
    a) A valid augmented bar chart; 
    b) An invalid augmented pie chart.}
    \label{fig:training}
\end{figure}

\cmo{
To maintain perceptual effectiveness,
we propose that the visual encodings in $V_s$ and $V_v$
should be consistent with respect to C3, \emph{i.e.}, the same visual values mean the same data.
Otherwise, the visualization design is invalid.
For example, \autoref{fig:training} shows that when augmenting a $V_s$ with additional data items, a bar chart visualization is \textit{valid}; a pie chart visualization is \textit{in}valid since the arc length of the virtual pie chart leads to inconsistent mappings (C3).}

\subsection{How can a static visualization be augmented by AR?}
\label{ssec:augmentations}

\cmo{As mentioned in the previous section, 
The design space of augmented static visualizations
can be determined by two dimensions: 
\textit{i)} the \textbf{visual encodings} of $V_s$ and $V_v$,
which can be \textbf{different} or the \textbf{same};
and \textit{ii)} the \textbf{composition} of $V_s$ and $V_v$,
which can be an \textbf{integrated} view or two \textbf{separate} views.}
We use these two dimensions to construct a design space (\autoref{fig:augmentations}), 
which outlines four ways to augment $V_s$:

\begin{figure}[h]
    \centering
    \includegraphics[width=0.99\columnwidth]{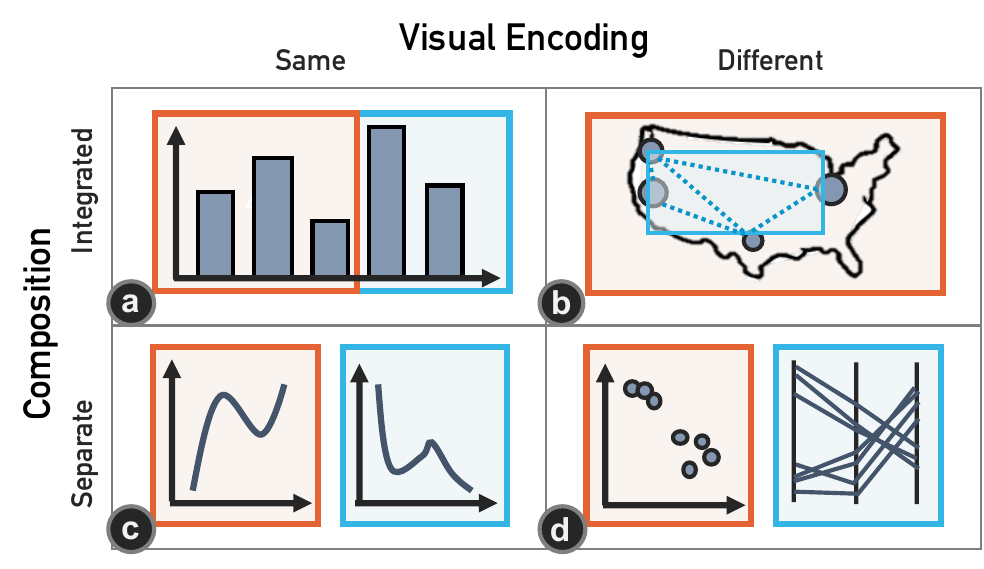}
    \caption{Four ways to augment static visualizations using AR: 
        a) Extended View;
        b) Composite View;
        c) Small Multiple;
        d) Multiple View.
    }
    \label{fig:augmentations}
    \vspace{-2mm}
\end{figure}

\begin{itemize}[noitemsep]
    \item \textbf{Extended View \cmo{\textit{(Same visual encoding  + Integrated composition)}}.}
    \autoref{fig:augmentations}a presents the scenario 
    that $V_v$ has the same visual encodings with $V_s$
    and is integrated with $V_s$ in a single view. 
    \textit{\underline{Augmentation Point}}:
    In this augmentation method,
    $V_s$ and $V_v$ are actually the same visualization (\emph{i.e.}, use the same visual encodings in the same view) but different in data.
    The designer can extend $V_s$ with additional data,
    allowing a static visualization 
    to present continually updated data 
    or to provide details on demand.
    \textit{\underline{Validity}}:
    A particular validation point of this method
    is that the designer should consider the data dependency between $V_s$ 
    and $V_v$.
    Not all $V_s$ can be augmented in this way.
    For example, 
    a pie chart 
    $V_s^{pie}$ 
    cannot be augmented with more data items through 
    $V_v^{pie}$.
    Specifically,
    $V_v^{pie}$ 
    can only match the ``updated'' 
    $V_s^{pie}$ 
    but not $V_s^{pie}$  
    itself.
    Thus, $V_v^{pie}$ will hinder the perception of $V_s^{pie}$.
    In other words, 
    using this method, the designer must ensure $V_s^{pie}$ will not change when the new data is appended.
    
    \item \textbf{Composite View \cmo{\textit{(Different visual encoding + Integrated composition)}}.}
    \autoref{fig:augmentations}b depicts the condition when 
    $V_v$ has different visual encodings with $V_s$ 
    and is integrated with $V_s$.
    \textit{\underline{Augmentation Point}}:
    When augmenting $V_s$ in this way,
    the designer can visualize another dataset using $V_v$
    that complements $V_s$,
    or can present extra data attributes of the dataset of $V_s$.
    \textit{\underline{Validity}}:
    Given $V_v$ 
    has different visual encodings from $V_s$,
    the designer does not need to concern the data dependency between
    them and should focus on general design issues, \emph{e.g.}, occlusions between $V_v$ and $V_s$.
    

    \item \textbf{Small Multiple \cmo{\textit{(Same visual encoding + Separate composition)}}.}
    \autoref{fig:augmentations}c 
    shows the scenario where $V_v$ has the same visual encodings with $V_s$ 
    and is displayed as a separate view.
    \textit{\underline{Augmentation Point}}:
    $V_v$ can be used to present other datasets in the same visual encodings of $V_s$.
    Generally, the result of this method is a small multiple.
    \textit{\underline{Validity}}:
    The augmentation will always be valid as
    $V_v$ is displayed separately from $V_s$.
    However, 
    the designers should consider whether the visual encodings fits for different datasets, \emph{i.e.}, the scalability issue.

    \item \textbf{Multiple Views \cmo{\textit{(Different visual encoding  + Separate composition)}}.}
    \autoref{fig:augmentations}d demonstrates the case 
    that $V_v$
    has different visual encodings with $V_s$
    and is displayed separately from $V_s$.
    \textit{\underline{Augmentation Point}}:
    Choosing this augmentation
    allows the designer to use AR to 
    extend $V_s$ 
    to a multiple view (\emph{e.g.}, a dashboard),
    presenting different perspective of the data behind $V_s$,
    or use $V_v$ 
    to visualize new datasets along with $V_s$ 
    that yields deeper insights into the data.
    \textit{\underline{Validity}}:
    This method can always ensure valid AR visualizations
    and has the most flexibility 
    as $V_s$ and $V_v$ are completely independent (both in data and visually).
    
\end{itemize}

In summary,
in our design space,
$V_s$ can be augmented in four different ways
with different augmentation points.
Among these four augmentation methods,
\textit{Extended View} is the most strict one,
as in which $V_s$ and $V_v$ are highly correlated;
\textit{Multiple View} is the most flexible one,
as in which $V_s$ and $V_v$
are completely independent of each other.

\subsection{Design Goals}
To create augmented static visualizations,
we conceive our authoring environment to accomplish three main goals:

\textbf{G1: Integrate the visualization design in one tool}---\cmo{To assure \textit{graphical consistency} (C1) and \textit{validity} (C3) between $V_s$ and $V_v$,
we aim to integrate
$V_s$ and $V_v$ into one single specification
and allow designers without AR expertise
to create both $V_s$ and $V_v$ in a single tool simultaneously:
specify $V_s$ and $V_v$, 
test and debug, and deploy.}

\textbf{G2: Preview the visualization design in one platform}---\cmo{To assure \emph{readability} (C2) of an augmented static visualization,
the designer may need
to work back-and-forth between the desktop platform and AR devices 
(\emph{e.g.}, head-mounted displays, mobile handheld).
This kind of cross-device workflow is tedious and time-consuming.
The authoring environment should alleviate this burden as much as possible.
Given that $V_s$ will be part of the reality,
we can preview the $V_v$ 
together with the $V_s$
on the desktop platform to simulate the AR scenario. Moreover, some augmentations (\emph{e.g.}, \textit{Composite View}) 
may display data that is unknown during the process of visualization design.
Therefore, an authoring environment must allow previewing the augmented static visualization.}

\textbf{G3: Provide automatic design support}---\cmo{The designer must assure \emph{validity} (C3) of an augmented static visualization.}
As indicated by our design space,
designing valid \textit{Extended View} is the most challenging
since in which there are data dependencies exist between the $V_s$ and $V_v$.
When new data is appended and visualized by the $V_v$,
a mismatch between the $V_s$ and $V_v$ can easily happen
due to inappropriate visual encodings,
leading to an invalid augmented static visualization.
When the data is large, or the visual encodings are complex, 
it will be difficult for the designer 
to verify and debug the inappropriate design manually.
Thus, the authoring environment should
automatically verify the design
and provide hints for debugging inappropriate visual encodings.

\section{Usage Scenario}
\label{sec:usage-scenario}
To illustrate how \tool{} accomplishes
the three design goals and introduce the full workflow,
we describe how Bob, a hypothetical visualization designer,
creates the node-link diagram in~\autoref{fig:gallery}a.

Bob is a teaching assistant coordinate of the Computer Science and Engineering department (CSE). 
He is asked to create a poster 
for introducing the CSE to the prospective post-graduate (PG) students.
He plans to use a hierarchical visualization to present the organization structure from the university to the department as an overview,
as well as the faculties in the CSE for details.
Since presenting the complete hierarchical structure requires a large area,
Bob has to hide some details 
in the poster.
Besides, some new faculties might join the CSE in the future,
making the current poster outdated.
Considering these issues, 
Bob decides to use \tool{} 
to create an augmented hierarchical visualization in the poster.

\begin{figure*}[t]
    \centering
    \includegraphics[width=1.99\columnwidth]{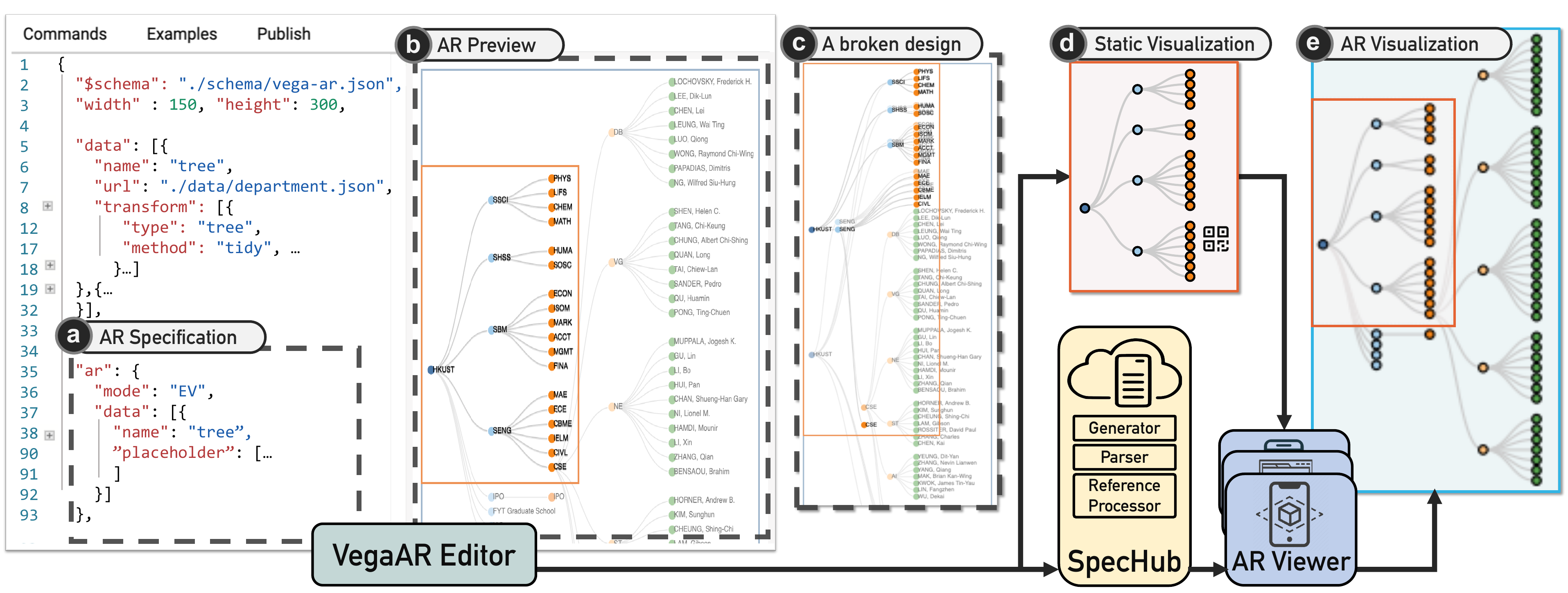}
    \caption{The main user interface and pipeline: 
        1) The designer specifies the static and virtual (in the \texttt{ar} block) visualization in the VegaAR Editor, which preview the result on the desktop platform.
        2) Then he publishes the specification to the SpecHub and exports the static part for printing. The SpecHub will further process the specification and prepare all prerequisites for the augmented static visualization.
        3) Finally, the audiences can use an AR viewer to
        scan the static visualization,
        fetch the virtual part from the SpecHub,
        and combine them to display the augmented static visualization.
    }
    \vspace{-2mm}
    \label{fig:pipeline}
\end{figure*}

Bob opens \tool{} and 
decides to choose a node-link tree example to initialize his design.
Bob loads the data in the Vega code
and defines the data to be displayed in AR in the \texttt{ar} block
(\autoref{fig:pipeline}a). 
For the uncertain child nodes (\emph{e.g.}, the new faculties),
he defines a placeholder for them in the \texttt{ar} block 
using wildcard characters.
\tool{} previews how
the tree visualization
looks like in AR (\autoref{fig:pipeline}c)
by showing both the real and virtual parts which are indicated
by the orange and blue border boxes, respectively.

In the preview (\autoref{fig:pipeline}c),
Bob notices a terrible mismatch between the real and virtual visualizations.
But he does not know why this happens. 
His attention is then attracted by a warning hint
showing in the Vega code (\autoref{fig:dataflow}a).
Reading this hint,
Bob notices that this mismatch is caused by the ``cluster'' layout, 
which places leaf nodes of the tree at the same depth.
Thus, when new virtual nodes with different depths 
are appended to the leaf nodes of the real tree,
all the nodes of the real tree need to be repositioned to
make sure that the added virtual leaf nodes
can be placed at the same depth with the real leaf nodes,
leading to the mismatch.
Following the hint,
he changes the layout from ``cluster'' to ``tidy'',
which appends new nodes without changing the existing nodes.
The mismatch problem is then solved (\autoref{fig:pipeline}b).

Satisfied with the design, Bob clicks the \texttt{publish}
button to export his design.
\tool{} automatically separates the real and virtual part
and generates the tree for presenting in reality with a QRCode that identifies
the tree diagram (\autoref{fig:pipeline}d).
For the virtual part,
\tool{} pushes the Vega specification to a backend server
for further processing,
including generating an AR reference image of the real tree so that it can be recognized by AR viewers,
converting the virtual part into AR environments,
and hosting it in a repository that can be fetched by AR viewers.
These details are handled by a ``black box'' that Bob does not need to concern.
Finally, Bob uses an AR viewer on his mobile phone to
scan the real visualization and 
observe the virtual visualization (\autoref{fig:pipeline}e).

\section{PAPARVIS DESIGNER}
\label{sec:tool}

This section describes the design and implementation of \tool{}, 
our authoring environment for augmented static visualizations.

\subsection{Workflow}

\begin{figure}[thb]
    \centering
    \includegraphics[width=0.95\columnwidth]{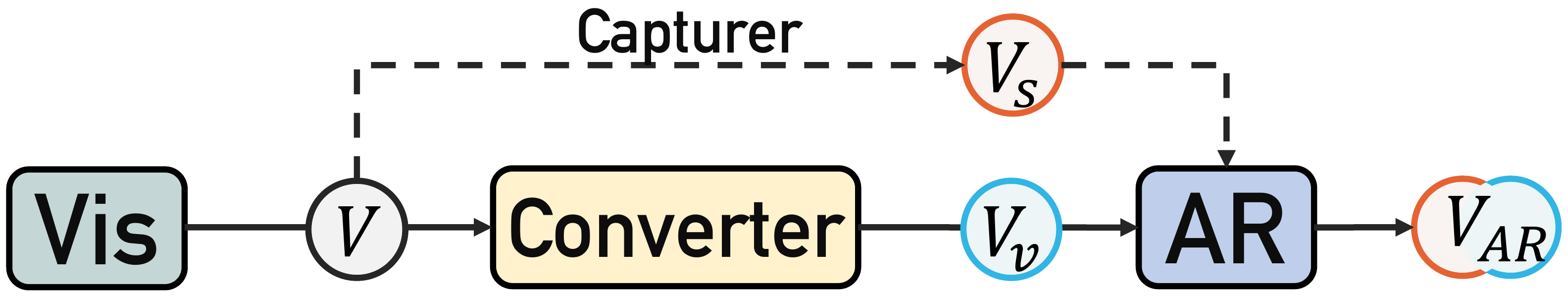}
    \caption{A workflow creates both static and virtual visualizations: 
     a ``stem'' visualization~\inlinefig{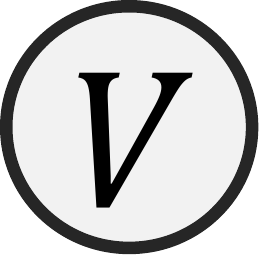} 
     can be rendered as a static visualization~\inlinefig{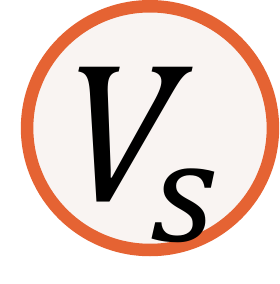}
     and a virtual visualization~\inlinefig{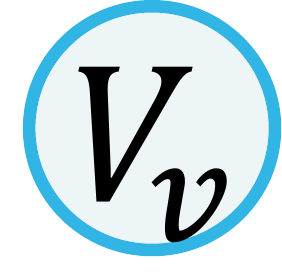},
     which are then combined as an augmented static visualization~\inlinefig{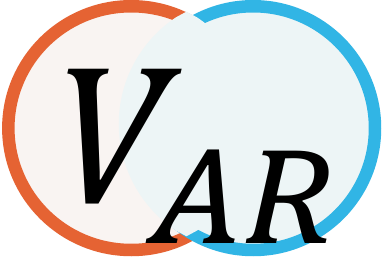}. 
     }
    \label{fig:workflow}
    \vspace{-8mm}
\end{figure}

\tool{} creates both static and virtual visualizations.
Instead of developing a tool that combines both
the functionalities of visualization tools
for desktop and for AR environments,
we abstract the common part of them
and propose a decoupled workflow (\autoref{fig:workflow}).
By separating the specification
(\emph{i.e.}, designing visualizations) 
from the deployment
(\emph{i.e.}, presenting visualizations),
our workflow allows designers without AR expertise 
to create both $V_s$ and $V_v$ (G1)
in \cmo{a single specification}
and leave the deployment to AR experts.

Specifically, 
our workflow adopts a browser/server architecture.
In the workflow, 
the designer uses a visualization authoring tool 
to create a ``stem'' visualization 
which can then be captured as a static visualization $V_s$. 
The designer can continue updating the ``stem'' visualization, 
similar to update a web page,
and use a converter 
to convert it into the virtual visualization $V_v$. 
An AR viewer 
can be used to 
combine the $V_s$ and $V_v$
to obtain the augmented static visualization. 
By this mean, both 
$V_s$ 
and 
$V_v$ 
are created based on the same specification,
thus maintaining a high level of consistency.

We implement the workflow in \tool{} (\autoref{fig:pipeline}),
which consists of three main components: 
\begin{itemize}[noitemsep]
    \item \textbf{VegaAR Editor} is built based on 
    a UI editor of Vega~\cite{vegaEditor}.
    In VegaAR Editor,
    the designer can create a $V_s$
    following the same practice in Vega
    and specify the $V_v$ in an \texttt{ar} block,
    which will be further introduced in the next section (Sec.\ref{ssec:preview}).
    When finishing the creation,
    the designer can \texttt{publish} the design
    to export the $V_s$ (\autoref{fig:pipeline}d)
    and push the whole specification, 
    including the \texttt{ar} block, to \textit{SpecHub}.
    VegaAR Editor will generate a QRCode that 
    links the $V_s$ to the specification on SpecHub.

    \item \textbf{SpecHub} is deployed on a cloud server 
    to receive and host the specifications from VegaAR Editor 
    \cmo{(\emph{i.e.}, like a piece of code hosted on Github)}.
    Besides,
    it prepares all prerequisites for the AR visualization,
    such as processing the specification of $V_s$ 
    to generate the AR reference image so that $V_s$ 
    can be recognized by AR viewers,
    parsing the \texttt{ar} block to render the $V_v$
    \cmo{with the new data whenever a user views it using an AR viewer.}
    
    \item \textbf{AR Viewer} is the endpoint (\emph{i.e.}, browser) 
    for audiences to view the augmented static visualization.
    When scanning a $V_s$,
    an AR viewer will identify the $V_v$ based on its QRCode, 
    fetch its corresponding $V_v$ from the SpecHub,
    and register $V_v$ onto $V_s$
    to present the augmented static visualization (\autoref{fig:pipeline}e).
    The AR viewer is not platform-specific.
    In this work, we have implemented three AR viewers
    on the iOS, Android, and web-based platform, respectively.
\end{itemize}

With this workflow, 
designers can create augmented static visualizations
in one tool without digging into the AR details.

\subsection{AR-Preview}
\label{ssec:preview}

We design \feata{} to avoid the frequently switching between devices (G2).
Specifically,
to preview $V_v$ together with $V_s$,
the designer must explicitly specify the $V_v$ during the creation,
which is not supported by the current Vega grammar.
Thus, we extend the Vega grammar
by adding an additional configuration block \texttt{ar} (\autoref{fig:pipeline}a) to it.
In the \texttt{ar} block,
the designer can explicitly specify the augmentation method,
the visual encodings,
and the data of a $V_v$,
no matter the data is certain or uncertain:
\begin{itemize}[noitemsep]
    \item \textbf{Certain}. 
    When the data is known
    (\emph{e.g.}, some detail information cannot be presented due to the space limitation),
    the designer can specify the $V_v$ 
    in the \texttt{ar} block,
    starting from defining its 
    augmentation method:
    1) in \textit{Extended View},
    the designer only needs to specify 
    new data and which datasets of the $V_s$ to be appended,
    and can reuse other visual encodings of the $V_s$;
    2) in \textit{Composite View}, 
    the designer can specify a new visualization in the \texttt{ar} block
    with new datasets or with the existing datasets of the $V_s$;
    3) in \textit{Small Multiple},
    the designer only need to specify the new datasets and the layout of the virtual small multiples in relation to the $V_s$;
    4) in \textit{Multiple View}, the designer is allowed to create a complete new visualization in the \texttt{ar} block 
    and define its placement with respect to the $V_s$.

    \item \textbf{Uncertain}.
    The data of the $V_v$ could be uncertain during the creation 
    (\emph{e.g.}, the data can only be obtained in the future).
    In this case,
    it will be arduous for the designer to imagine the $V_v$.
    We provide a \texttt{placeholder} mechanism to
    allow designers to generate mockup data using wildcard.
    In the \texttt{ar} block, 
    the designer can create placeholder datasets
    and generate their data by 
    specifying the data type (\emph{e.g.}, categorical, temporal, and quantitative),
    number of datum,
    and ranges or potential options of the values
    of each column.
\end{itemize}

\cmo{The physical space of the $V_v$ is determined by the
Vega specification (\emph{e.g.}, set the canvas width to 500 unit), relative to the $V_s$.}
By explicitly defining the $V_v$ in the \texttt{ar} block,
VegaAR Editor will provide visual preview of the virtual content 
together with the $V_s$,
thus reducing the switching between devices.

\subsection{Validator}

An \textit{Extended View} (see design space) can easily be invalid
given the data dependency between $V_s$ and $V_v$ (\emph{e.g.}, the pie chart in Sec.\ref{ssec:model} and the tree diagram in Sec.\ref{sec:usage-scenario}).
When choosing this augmentation method,
\tool{} automatically verifies the dataflow of the visual design
and provides hints for debugging invalid visual encodings (G3).

\begin{figure}[thb]
    \centering
    \includegraphics[width=0.99\columnwidth]{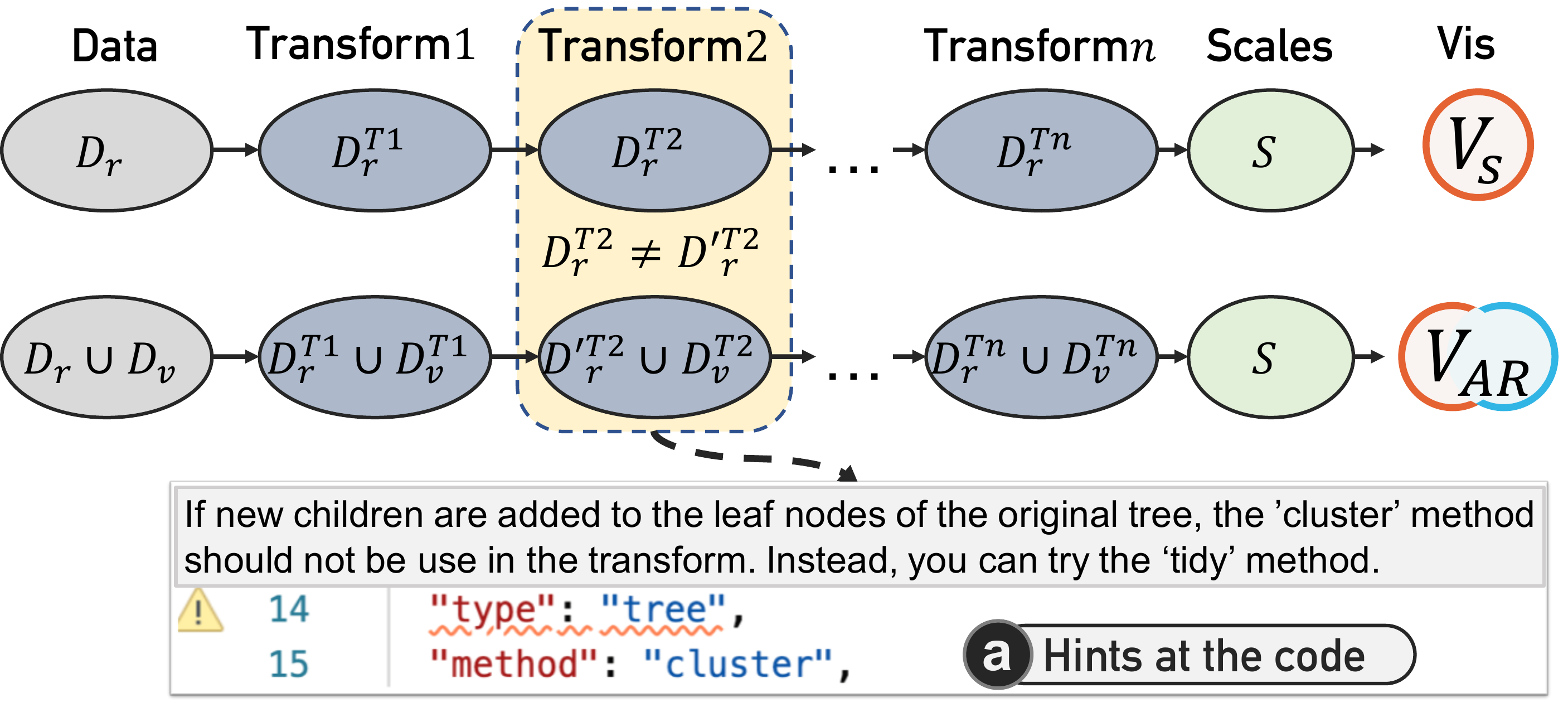}
    \caption{Validate the visual design by comparing the dataflows, thus enabling providing debug messages specific to a transform.}
    \label{fig:dataflow}
\end{figure}

Instead of using a rule-based method,
we propose to validate the visualization
by comparing the dataflow between a $V_s$ and an augmented static visualization.
A general visualization process of a dataset $D_r$
can be summarized in~\autoref{fig:dataflow}:
$D_r$ will first be processed by
a serials of data transforms
and then mapped to visual channels
to construct the $V_s$.
As demonstrated in Sec.\ref{ssec:augmentations},
in a valid augmented static visualization,
the $V_s$ should remain unchanged
after appending the new data $D_v$,
which means its all intermediate states $D_r^{T}$
before the visual mapping
should also be unchanged.
Thus,
for each data transform,
we can compare 
the intermediate states of the $V_s$ 
and the augmented static visualization
to see whether the intermediate states derived from $D_r$ changes.
If an intermediate state changes,
for example ${D'}_r^{T2} \neq D_r^{T2}$,
it means the transform (\emph{i.e.}, Transform$2$)
alters the $V_r$ after appending new data,
thus leading to an invalid augmented static visualization.
Therefore, 
we can give the designer a hint 
that is specific to a transform 
and further provide messages for fixing the invalid visualization (\emph{e.g.}, \autoref{fig:dataflow}a)
regardless of the complexity of the visualization. 
For example, 
when trying to augment a treemap, 
the system warns \texttt{avoid 'treemap' when new nodes are added to the internal nodes} 
as this would update the layout of the underlying treemap.

\subsection{Implementation}
The implementation of \tool{} mainly consists
of three parts, namely, VegaAR Editor, 
SpecHub, 
and AR viewers.
VegaAR Editor is a web-based application
that extends the Vega Editor~\cite{vegaEditor}, 
Vega Schema~\cite{vegaSchema} (\emph{i.e.}, for the \texttt{ar} block),
and Vega Compiler~\cite{vegaCompiler} (\emph{i.e.}, for the \feata{} and \featb{}).
SpecHub is a NodeJs-based application
that runs on a cloud server.
SpecHub also reuses the extended Vega Compiler in VegaAR Editor to parse
and generate $V_v$.
We have implemented AR viewers based on Vuforia~\cite{vuforia}
on three different platforms,
including iOS (iPhone8 Plus), Android (Huawei P10), and web-based platform.
All AR viewers only provide minimal functionalities,
including QRCode decoder, 
AR image recognition,
and 3D registration.

\begin{figure*}[h!]
    \centering
    \includegraphics[width=2.1\columnwidth]{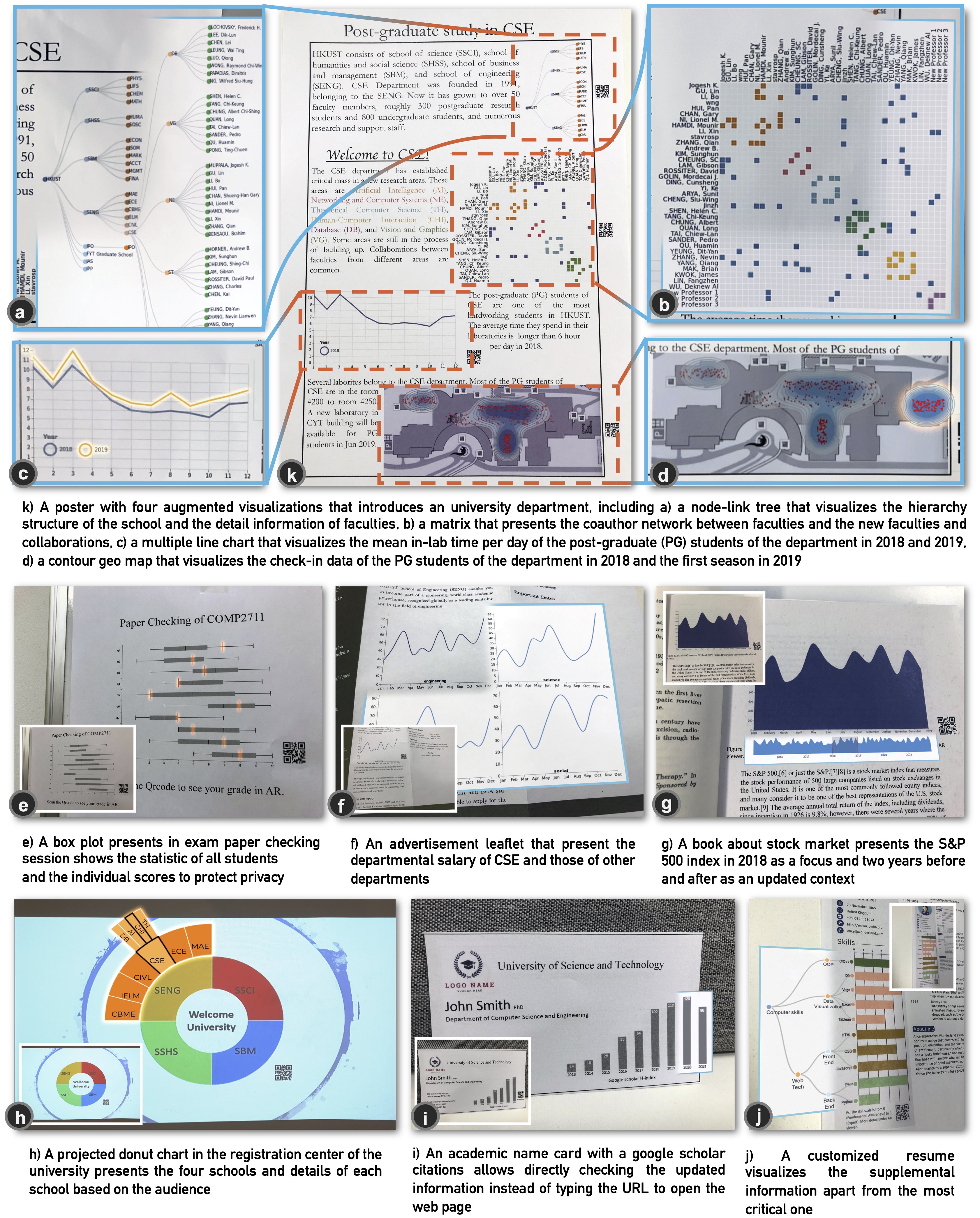}
    \caption{Examples created by \tool{}. 
    a) - d) show the augmenting results of k).
    In e) - j), the sub-figures show the static visualizations.
    }
    \label{fig:gallery}
\end{figure*}

\section{Examples and Scenarios}

\autoref{fig:outdated-tl} and \autoref{fig:gallery}
present examples varying in
data type,
chart type,
augmentation method,
media,
and purpose.
We summarize scenarios that can benefit from augmenting static visualizations.

\textbf{Overcoming space limitations}---AR can extend a static visualization where the space is scarce.
For example, 
in \autoref{fig:outdated-tl}a, 
given that much space of the newspaper is occupied by the banner,
the data journalism uses an augmented static visualization
to extend the canvas for visualization, thereby presenting additional data.
A similar usage can be found in the example of poster (\autoref{fig:gallery}a),
leaflet (\autoref{fig:gallery}f), name card (\autoref{fig:gallery}i),
and resume (\autoref{fig:gallery}j).
Moreover, 
in the wall-sized timeline in \autoref{fig:outdated-tl}b,
it is difficult to physically extend the wall for presenting updated data.
By using augmented static visualizations,
the designer can easily break through the limitation of physical world and present more information.

\textbf{Displaying new data}---AR can update the static visualizations,
thus providing the latest information 
and saving the cost to recreate them.
In real world,
information can be updated frequently
while the materials to present the information might not.
Augmented static visualization is a promising way to relieve
the mismatch between the update cycle of information and materials.
\autoref{fig:outdated-tl}b presents a representative example of using AR
to update a wall-sized timeline, which has been outdated for more than three years.
\autoref{fig:gallery}g demonstrates 
a stock market book
uses an augmented static visualization 
to provide the latest S\&P index data
to help the reader better understand the historical context.
Generally users can also use augmented static visualizations
to keep their information in static visualizations online.
For example, \autoref{fig:gallery}i is an academic name card
with google scholar citations that will not be outdated after distributing.

\textbf{Showing details}---AR can empower static visualizations with the capability to show details on demand.
``Overview first, zoom and filter, then details on demand''~\cite{Shneiderman1996}
is a well-known information visualization guideline
yet can hardly be followed in static visualizations.
Augmented static visualizations can achieve this classic technique on static visualizations.
For example, \autoref{fig:gallery}a presents 
the overview hierarchical architecture of a university
and collapses the detail nodes into AR, thus avoiding information overload. 
As shown in \autoref{fig:gallery}h and j, many tree diagrams can benefit from this feature.

\textbf{Complementing additional data}---For visualizations in public spaces,
augmented static visualizations 
can be used to overlay personal data.
For example,
in \autoref{fig:outdated-tl}c,
by overlaying the trajectories data onto a public map, 
the augmented static visualization complements the geographical visualization with the user's own data, thus helping the user to understand his/her movement pattern.
Another example of complementing visualizations is
to provide supplemental information.
The leaflet advertisement in \autoref{fig:gallery}f
shows the salaries of other departments in AR for comparison purpose,
since this extra information is not the focus of the advertisement 
but can help readers obtain the context information.

\textbf{Protecting privacy}---Augmented static visualizations 
can protect individuals privacy. In \autoref{fig:gallery}e, the teacher printed the class' summary statistics for an exam, overlaying a student's personal ranking in the augmented static visualization.
Similarly, \autoref{fig:gallery}h shows a projected chart for the new students to check their private information.

\vspace{-1em}
\section{User Study}
We conducted a controlled user study to assess the usability and the utility of \tool{}. The study aimed to evaluate whether visualization designers without AR expertise could create augmented static visualizations using \tool{} and whether the two advanced features (\feata{} and \featb{}) facilitated the creating process.


\newcommand{\base}{\textsc{Base}}
\newcommand{\pro}{\textsc{Pro}}
\newcommand{\validityTree}{\textit{Validity-tree}}
\newcommand{\validityMatrix}{\textit{Validity-matrix}}
\newcommand{\occlusion}{\textit{Occlusion}}
\newcommand{\unnoticeable}{\textit{Unnoticeable}}

\textbf{Baseline technique:}
As no comparable tool exists for a fair baseline comparison,
we provided two versions of \tool{}:
a \textit{base mode} (\base{}) and a \textit{pro mode} (\pro{}). \pro{} provided full features of \tool{}
while the \base{} did not provide the \feata{} and \featb{}. To ensure a fair comparison, \base{} also provided the \texttt{ar} block (without the visual preview)
for explicitly defining the virtual visualizations.

\textbf{Tasks:}
The study simulated creating both static and virtual visualizations. We took the poster example in \autoref{fig:gallery}k (without the visualizations)
and asked participants to complete the poster
by creating the four visualizations (referred to as tasks T1-T4). 
T1-T4 covered a wide range of data types
and were required to extend the visualization using an \textit{Extended View}, 
as we considered this extension is the most challenging and hence expected strongest evidence for our tool. 
For each task,
participants were provided with
the background information,
relevant Vega documents,
an instruction to load the dataset,
and one or two Vega examples to initialize the design.

\begin{enumerate}[label={\textbf{T\arabic*}}, leftmargin=*,noitemsep]
    \item \textbf{\validityTree{}} required participants to visualize a hierarchical dataset about the university structure. The provided examples were a node-link tree and a treemap.
    The treemap was ensured to be invalid for the dataset
    while the node-link tree was invalid in default but can be fixed to be valid by modifying \textit{one} visual encoding.
    The purpose of this setting was to assess 
    whether the participant can choose the proper example and debug the design in different modes.
    
    \item \textbf{\validityMatrix{}} had the same purpose as \textbf{T1} but required participants to visualize a network dataset about the collaboration among faculties in the college.
    The examples provided a matrix diagram and a radial node-link graph.
    The radial node-link graph was ensured to be invalid for the dataset while the matrix diagrams was invalid in default but can be fixed by modifying \textit{one} visual encoding.

    \item \textbf{\occlusion{}} required visualizing a temporal dataset about the mean in-lab time of students.
    The provided examples included a single line chart and a multiple line chart.
    Both examples were valid without any corrections.
    We deliberately put the text description of \textbf{T3} on the right 
    to see how the participants deal with the potential occlusions due to the AR extension in different modes.

    \item \textbf{\unnoticeable{}} required the participants to visualize a geographic dataset.
    The provided example was a contour map.
    The example was invalid in default for the dataset 
    but can be corrected by modifying \textit{one} visual encoding. 
    However, 
    the default invalid design 
    only led to an unnoticeable mismatch, 
    \emph{i.e.}, the mismatch of \textbf{contour}s,
    between the $V_s$ and $V_v$.
    The mismatch is expected to be detected by \featb{} and we want to see how the participants handle the conflict
    between their observation and our hints.
\end{enumerate}
The four tasks were divided into two groups (\validityTree{} \& \occlusion{} \emph{vs}. \validityMatrix{} \& \unnoticeable{}) to balance the workload in different modes.
The materials we provided to the participants can be found in the supplemental materials.

\textbf{Participants and Apparatus:}
We recruited 12 participants (8 male; age: 22-30, average 25.6),
who had at least two-year experience with Vega or D3 (since Vega is built based on D3)
and no expertise in AR programming (\emph{e.g.}, Unity).
According to the pre-study survey,
all participants had more than two years experience on data visualization ($\mu = 3.54, \sigma = 1.44$).
Each participant received a gift card worth \$14 
at the beginning of the session, independent of their performance. 
The study was run in the lab, using a 15-inch laptop, 
or their own.
A 5.5-inch iPhone8 Plus was provided to view the augmented static visualization.

\begin{figure}[th]
    \vspace{-2mm}
    \centering
    \includegraphics[width=0.99\columnwidth]{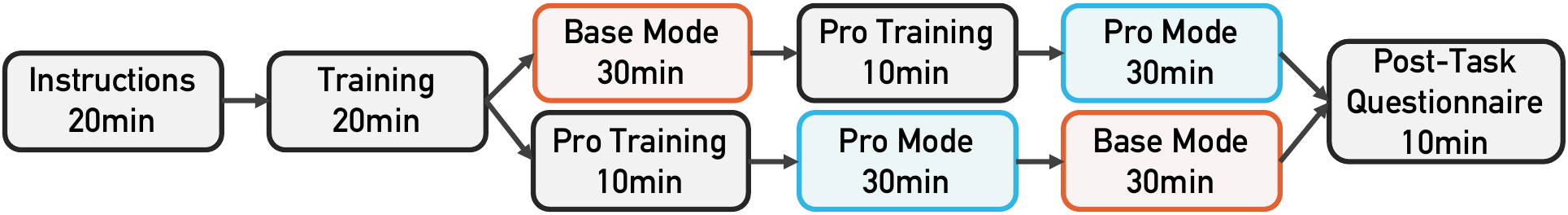}
    \caption{Participants completed four tasks, two in each of two conditions (\base or \pro). The conditions were counter-balanced across participants, thus each participant completed only one path in this figure.}
    \label{fig:userstudy-protocol}
    \vspace{-2mm}
\end{figure}

\textbf{Procedure:}
The procedure of our study is summarised in \autoref{fig:userstudy-protocol}.
We first introduced the purpose of the study,
explained the concept of \emph{Validity}
via examples (\emph{i.e.}, \autoref{fig:training}), 
introduced Vega, 
as well as how to create a visualization with \tool{} (\textit{introduction}, 20min).
Participants then were given a step-by-step \textit{training} (20min) instruction 
to reproduce the two examples in~\autoref{fig:training}.
Participants were encouraged to ask questions 
and explore examples.

We performed a within-subject study on the four tasks,
two in each condition: \base{} (30min) and \pro{} (30min).
We counterbalanced the order of the conditions across participants:
the \pro{}-group started with the \pro{} and then performed on \base{}, 
the \base{}-group did start with \base{}. 
Since a task cannot be done in both conditions, 
we also counterbalanced the order of the groups of tasks across participants.
Before performing with \pro{}, 
participants were given an instruction sheet of the two advanced features (\feata{} and \featb{})
and were encouraged to explore the \pro{} and ask questions (\textit{\pro{} training}, 10min).
Before each task,
participants were provided with the materials and
encouraged to 
ask question about and get familiar with the materials.
A task was started when a participant was confident and confirmed to begin
and ended when the participant confirmed finishing.
Participants finished with a \textit{post-study questionnaire} (10min, \cmo{adapted from~\cite{lund2001measuring}}) in which they rated their self-perceived efficiency, effectiveness, and mental demand on a scale from 1 (``better in \base{}'') to 7 (``better in \pro{}''). To collect subjective feedback for \pro{},
participants rated its usability, usefulness, and satisfaction and answered questions in a semi-structured interview. Each session lasted approximately 1.5-2 hours.

\textbf{Task Performance Measures:}
We recorded task completion \textit{time} 
and \textit{correctness} for each task.
A correct design 
a) does not occlude its text description
and b) is valid.
All measures were explicitly explained to the participants before the tasks.


\subsection{Quantitative Results}

\begin{figure}[th]
    \vspace{-2mm}
    \centering
    \includegraphics[width=0.99\columnwidth]{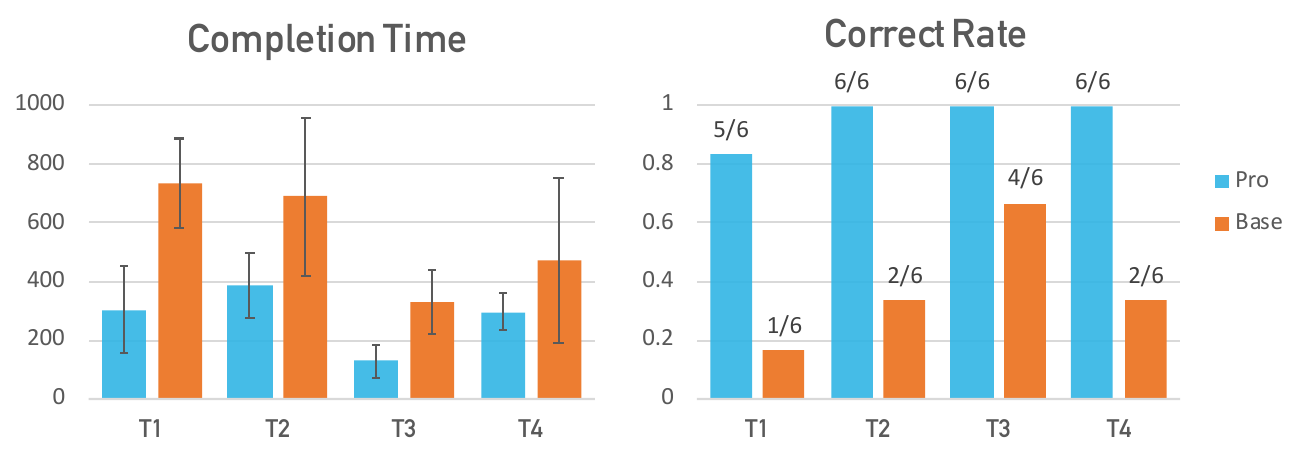}
    \caption{Left: Average task completion time and 95\% CIs across tasks. Right: The correct rate across tasks.}
    \label{fig:quant-results}
    \vspace{-2mm}
\end{figure}

\autoref{fig:quant-results} shows the results with 95\% confidence intervals (CIs). Significance values are reported for $p <.05 (\ast)$, $p <.01 (\ast\ast)$, and
$p <.001 (\ast\ast\ast)$, abbreviated by the number of stars.

\textbf{Completion Time:}
Participants finished the tasks faster with \pro{} ($\mu=278s$, $95 \%CI = [206, 351]$)
than with \base{} ($\mu = 557s$, $95 \%CI = [413, 701]$).
Before more detailed examinations,
we first confirmed that all results of completion times in each condition
follow normal distribution using Anderson-Darling test.
Using an independent-samples t-test with a null hypothesis that
the participants perform equally fast in each mode,
we found that participants performed significantly
faster in \pro{} on \validityTree{} ($\ast\ast$), \validityMatrix{} ($\ast$), 
and \occlusion{} ($\ast\ast$). 
No significant difference was observed for completion time with \unnoticeable{} ($p = 0.14$).

\textbf{Correctness}:
Almost all the augmented static visualizations (23/24) created in \pro{} were correct
while far less were correct with \base{} (9/24). Correctness follows a Bernoulli distribution (\emph{i.e.}, 1 \emph{vs.} 0),
and we used Binomial test with a null hypothesis 
that an augmented static visualization has a equal chance to be correct or wrong,
by which significant positive effects were observed
on \validityMatrix{} ($\ast$), \occlusion{} ($\ast$), \unnoticeable{} 
($\ast$) in \pro{},
and no significant effect was observed in others. 

\begin{figure}[th]
    \centering
    \includegraphics[width=0.99\columnwidth]{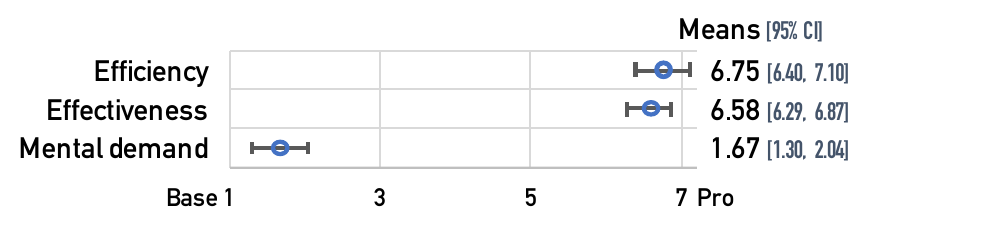}
    \caption{Means and 95\% CIs of post-study self-reported measures on a scale from 1 (better in \base{}) to 7 (better in \pro{}). \cmo{Mental demand accesses how much mental activity was required.}}
    \label{fig:self-report}
    \vspace{-2mm}
\end{figure}

\textbf{Post-Study Feedback:}
Anderson-Darling test reveals that the post-study self-reported results (\autoref{fig:self-report})
did not follow normal distribution. 
Thus, we used a one-sample non-parametric Wilcoxon signed rank test 
with a null hypothesis that the result is middle Likert scale value. 
We found significant positive effects for participants' reported efficiency ($\ast\ast\ast$) and effectiveness ($\ast\ast$) in \pro{}. 
For \base,
a negative effect was observed for
the participants' self-reported mental demand ($\ast\ast$).

\subsection{Qualitative Results}

\begin{figure}[th]
    \centering
    \includegraphics[width=0.99\columnwidth]{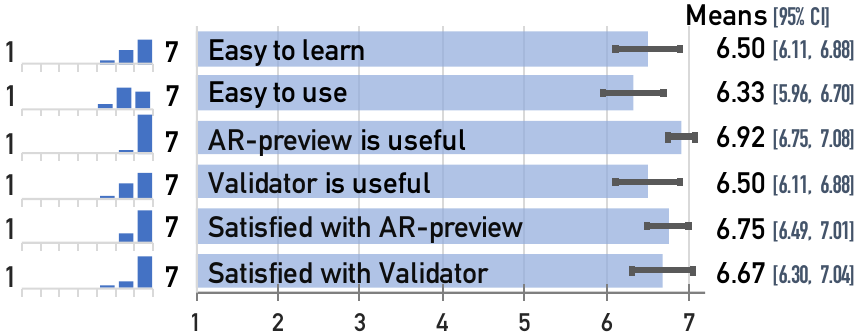}
    \caption{Subjective feedback on the usability, showing means and 95\% CIs. Distributions are shown on the left.}
    \label{fig:subjective}
    \vspace{-.5em}
\end{figure}

\textbf{Usability:} Generally, participants lauded the usability of our system as \textit{easy to learn} ($\mu=6.50, 95\% CI = [6.11, 6.88]$) and \textit{easy to use} ($\mu=6.33, 95\% CI = [5.96, 6.70]$) (Figure \ref{fig:subjective}).
P8 commented that \textit{``I have never tried Unity and AR things''} but \tool{} \textit{``can help me quickly produce AR extensible visualizations.''}
When we asked the participants 
whether they want more control of the AR details
(\emph{e.g.}, visual effects),
participants preferred that the tool would be cost-effective and remained simple.

\vspace{-1em}

\textbf{Usefulness:}
As for the two features of our authoring environment,
the participants commented positively and confirmed
the usefulness of \feata{} ($\mu=6.92,~95\%~CI = [6.75, 7.08]$)
and \featb{} ($\mu=6.50,~95\%~CI = [6.11, 6.88]$).
Although we provided an AR viewer for participants
to check their design in the study,
participants didn't use it frequently since
\textit{``\feata{} is enough thus no need to switch to the AR device''} (P2)
and \textit{``it provides immediate feedback.''} (P9).
Participants also agreed the usefulness of \featb{}.
As pointed out by P4,
who had 6 years experiences on data visualizations,
\textit{``the \featb{} is really important for the debug''}.
Indeed, the participants were able to finish
the creation more quickly and effectively with these two features,
just like P10 noted that \textit{``creation process is tedious without \featb{} and \feata{}.''}

\textbf{Satisfaction:}
The participants responded with a high user satisfaction for \feata{} ($\mu=6.75,~95\%~CI = [6.49, 7.01]$)
and \featb{} ($\mu=6.67,~95\%~CI = [6.30, 7.04]$). 
They said \feata{} was \textit{``intuitive''}.
Other comments mentioned improvements for \featb{},
including \textit{``better highlight the problematic code''} (P2)
and \textit{``improve the error descriptions''} (P7).

\section{Findings Summary}
Our study shows that all participants were able 
to create augmented static visualizations using \tool{}.
We discuss further observations and insights here.

\textbf{Whose faults? AR or my design?}---As discussed in Sec.~\ref{ssec:model},
\textit{in}appropriate visual encoding may lead to a
mismatch between the static and virtual visualizations, \emph{e.g.},
the color, sizes, and positions.
During the study, 
we observed that some participants suspected 
that the AR viewer cannot correctly align the $V_v$
with $V_s$
at the first time when there was a misalignment happened between the $V_s$ and $V_v$.
Even after confirming with us that there was nothing wrong with the AR part, 
some participants still doubted 
that the AR device worked properly,
especially when designing without the help of \featb{}.
In practice, 
given that the AR details are similar to a ``black box'' in our workflow,
it is important to help designers distinguish the error from AR and their designs.

\textbf{Where am I? Reality or virtuality?}---Creating augmented visualizations requires designers
to maintain two designs for $V_s$ and $V_v$ in mind.
Although 
\feata{} can alleviate this burden,
some participants 
could not consider both $V_s$ and $V_v$ simultaneously in the tasks. 
For example,
some participants chose the single line chart to present the temporal data in \occlusion{}.
When they discovered that
the virtual part would occlude the text description,
they attempted to flip the $V_s$ to a right-to-left orientation
and then struggled in this counter-intuitive design.
An interesting fact was that 
all the participants 
who 
chose the single line chart and 
struggled in this issue
had more than 3 years experiences in visualization design
while those with less than 3 years experiences directly
chose the multiple line chart.
We suspected this was due to more experienced participants having more clear routines.
How to effectively help designers closing the gap between
reality and virtuality during creation is one of the important future improvements.
\cmo{Allowing the designer to create the visualization design
in-situ rather than on a desktop can be a potential solution.}

\textbf{What is the hint? Ignore or follow?}---In \pro{},
our authoring environment provides useful hints 
for debugging augmented static visualizations.
In the study, we observed that one participant,
the only one who failed in \validityTree{} in \pro{},
ignored any hints. 
In the interview, 
he said \textit{``I thought the hints are useless so I didn't read it [sic].''}
Yet, the majority of participants followed the hints but
admitted that they actually did not know what was wrong with their designs.
We also observed that the hints occasionally confused the participants.
For instance, in \unnoticeable{},
the \textit{contour}s between the $V_s$ and $V_v$ were not consistent,
which was too subtle to be noticed by the participants 
but can be detected by \featb{}.
Thus, participants got confused by the hints 
and spent relative long time on \unnoticeable{},
which we believe was the reason of no significant difference between \base{} and \pro{} was observed on \unnoticeable{}.
We discussed this issue with the participants;
11 out of 12 agreed that the hints from \featb{} seemed as a ``strict mode warning''
instead of an error.

\cmo{\textbf{What scenarios can augmented static visualizations be used for?}---We were interested in what kind of scenarios the participants could envision augmented static visualizations. 
In the interviews, all participants suggested that such kind of visualizations can be used for 
public display, such as information boards or park maps, 
and interactive artworks (P4-P7, P11) in exhibitions or as gifts (P2, P5).
}

\section{Future work and Limitations}

\textbf{Generalization to 3D and dynamic AR visualizations}---\tool{} 
currently focuses on 2D static AR visualizations
but it can be generalized to 3D and dynamic AR visualizations.
As has been proven successful in DXR~\cite{Sicat2019},
Vega grammar can be extended from 2D to 3D visualizations.
It is possible to adopt the extension and design from DXR,
thus allowing designers to create 3D $V_v$ in \tool{} \cmo{(examples can be found in https://github.com/PapARVis)}.
Meanwhile, the Vega grammar inherently supports interactive visualizations,
providing opportunities to bring interactivity and animation to static visualizations.

\textbf{Multiple augmentations for collaborations}---Our design so far concentrates on augmenting one $V_s$ on one AR device.
A promising but more challenging question is
how to design AR visualizations distributed 
across multiple devices to support collaborations.
\cmo{Challenges, for example, include 
how to design cross-device interactions 
(\emph{e.g.}, HMDs prefer mid-air gestures and handheld devices use touch), or
how to support collaborations without sacrificing privacy protection.
}


\textbf{Design with scene understandings}---
In many real-world examples, the immediate environment influences readability and how a visualization is read. 
For example,
designers can create $V_v$ that 
adapts to the ambient light,
visualizes in-situ temperature data, 
or adapt more deeply to the AR-Canvas~\cite{bach2017drawing}. 
Recent progress in computer vision on visualizations~\cite{chen2019a}
offers possibilities for environment adaptive visualizations.

\textbf{Study Limitations}---Similar to other studies of authoring tools~\cite{Amini2017, Kim2019, Xia2018}, the sample size of our user study is small, given that access to experts is naturally limited. Further evaluation is thus suggested.
\tool{} currently generates a QRCode along with a $V_s$
when publishing, which size and position can be configured in the \texttt{ar} block.
While the QRCode has a good usability and is a flag to attract audiences to scan, it adds a superfluous component to the visualization. Future study is required to optimize the QRCode
by striking a balance between functionality and aesthetics.

\section{Conclusion}
In this paper we presented and evaluated \tool{}, an authoring environment for augmenting static visualization with virtual content. It integrates the workflow of creating both static and virtual visualizations by extending Vega, a popular visualization grammar.
Two features are provided to facilitate the creation process:
\feata{} reduces the switching between platforms and \featb{} provides debugging hints.
We provided an exemplary gallery to demonstrate the expressiveness of \tool{} as well as the broad application prospect of augmented static visualizations.
A user study showed that
\tool{} enabled visualization designers with no AR development experiences to create augmented static visualizations.
We have also shared and discussed the insights from our study, which implies future research.
\section{Acknowledgements}
This project is partially supported by a grant from the Research Grants Council of the Hong Kong Special Administrative Region, China (Project No. AoE/E-603/18).

\newpage

\balance{}

\bibliographystyle{SIGCHI-Reference-Format}
\bibliography{sample}

\end{document}